\documentclass[11pt,twoside]{article}

\usepackage{asp2004}
\usepackage{epsf}
\usepackage{lscape}

\newcommand{\HII}{H\,{\sc ii}}
\newcommand{\Fe}{$^{60}$Fe}
\newcommand{\Be}{$^{10}$Be}
\newcommand{\Al}{$^{26}$Al}

\begin{document}
\title{Understanding our Origins:  Star Formation in \HII\ Region Environments}   
\author{J. Jeff Hester \& Steven J. Desch}   
\affil{Arizona State University, Department of Physics \& Astronomy, Tempe, AZ 85287-1504, USA}    

\begin{abstract} 
Recent analysis of the decay products of short-lived radiounclides
(SLRs) in meteorites, in particular the confirmation of the presence
of live \Fe\ in the early Solar System, provides unambiguous evidence
that the Sun and Solar System formed near a massive star.  We consider
the question of the formation of low-mass stars in the environments
of massive stars, presenting a scenario for the evolution of a star
and its disk as it is overrun by the ionization front at the edge of
an expanding \HII\ region.  The stages in this scenario include: (1)
compression of molecular gas around the edge of the \HII\ region; (2)
induced low-mass star formation in this compressed gas; (3) an ``EGG''
phase when a dense star-forming clump is overrun by the ionization
front; (4) a ``proplyd'' phase during which the disk is truncated by
photoevaporation; (5) a long-lasting phase during which a young star
and its truncated disk evolve in the hot, tenuous interior of an \HII\
region; and (6) a phase when the ejecta from one or more nearby supernova
explosions overruns the disk, injecting SLRs including \Al\ and \Fe.
Most of these stages can be observed directly.  The exceptions are stage
(2), which must be inferred from the localization of low-mass protostars
in compressed molecular gas near ionization fronts, and stage (6), which
is an unavoidable consequence of the presence of low-mass protostars seen
near massive stars that will go supernova within a few million years.
(This differs from models in which the same supernova is responsible
for both triggering the formation of a star and injecting SLRs.)
This mode of star formation may be more characteristic of how most
low-mass stars form than is the mode of star formation seen in regions
such as the Taurus-Auriga molecular cloud.  We discuss the implications
of this scenario for our understanding of star formation, including the
possible role of photoionization in limiting the masses of stars.  We also
discuss the implications of the young Sun's astrophysical environment for
our understanding of the formation and evolution of the Solar System.
These include the effects of intense UV radiation from nearby massive
stars on the structure and chemistry of the disk, dynamical effects
due to close encounters of the Solar System with other cluster members,
and the role of the decay of SLRs in the evolution of the Solar System.
We conclude that low-mass stars and their accompanying disks form and
evolve differently near massive stars than they do in regions like
Taurus-Auriga, and that these differences have profound implications
for our understanding of our origins.
\end{abstract}



\section{Introduction}

Two of the great scientific success stories of the last several decades
are our growing understanding of the way stars form, and our ability
to reconstruct the history of our own Solar System.  These two lines
of scientific investigation meet in the Sun's protoplanetary disk.
To date we have only begun the task of carefully merging theories of
star formation with theories of the early evolution of the Solar System.
That task is one of the central challenges for both fields in the years
to come.  Given the extraordinary amount of information and ideas that
have emerged from those investigations, we should not be surprised if our
theories of both Solar System formation and evolution and our theories of
the formation of low-mass, Sun-like stars will each need to be modified
on the basis of insights gained from the other.

\begin{figure}[!ht]
\plotone{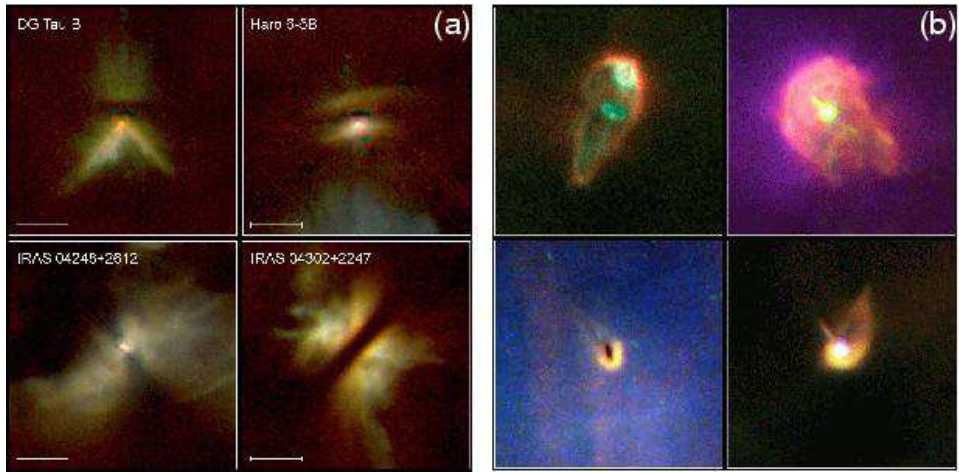}
\caption{(a) {\it Hubble Space Telescope} images of four young stellar
objects and their surrounding disks located in the Taurus-Auriga
molecular cloud \citep{pad99}.  (b) {\it Hubble Space Telescope} images
of four young stellar objects (proplyds) with photoevaporating disks,
located in the interior of the Orion Nebula \HII\ region \citep{bom00}.
}
\label{YSODISKS}
\end{figure}

Low-mass stars form in a variety of different environments.  
Figure~\ref{YSODISKS}a shows {\it HST} images of a number of young stellar
objects forming in relative isolation from each other in the nearby
Taurus-Auriga molecular cloud.  These protostars sit in the
centers of large dusty disks buried in the cold dark molecular cloud.
In contrast, Figure~\ref{YSODISKS}b shows {\it HST} images of a number
of young stars and protoplanetary disks sitting in the hot, ionized
interior of the Orion Nebula.  This region has formed a rich cluster of
stars with a number of luminous high-mass stars at its center.  The disks
here are being photoevaporated by intense ultraviolet radiation from those
nearby massive stars.  It takes little more than
a glance comparing Figures~\ref{YSODISKS}a and \ref{YSODISKS}b to realize that
the conditions under which stars form and disks evolve can be radically
different, depending on their astrophysical setting.

Meteorites play a crucial role in efforts to understand the history of the
Sun and Solar System. Meteorites provide detailed information about the
chemical, physical, and thermodynamic properties at different locations
within the disk, as well as adding constraints on the time scales
and physical processes by which objects in the Solar System formed.
In particular, the presence of the decay products of short-lived
radionuclides (SLRs) in meteorites offers hope of establishing the
type of astrophysical environment in which the Sun and Solar System
formed.  These radionuclides, such as \Al, have half-lives of only a
few million years or less.  While most of the chemical elements in the
Solar System more massive than hydrogen and helium were synthesized in
stars that lived and died over the course of the 9 billion years that
transpired before the Sun formed, SLRs have such short lifetimes that
they must have instead originated more locally, close to the time of
the Sun's birth.  SLRs might have been formed by spallation reactions
in the Sun's protoplanetary disk due to cosmic rays from solar flares,
or they might have been synthesized in nearby stars and then injected
into the formation environment of the Sun by supernovae or other means.
Establishing which origin is correct provides the key needed to place
the birth of the Sun and Solar System in its proper astrophysical context.

In this paper we propose an answer to that question.  The presence
of live \Fe\ in the early Solar System can only be understood if the
Sun formed in close proximity to a massive star that went supernova.
Building on our earlier work \citep[e.g.,][]{hes96,hes04} we consider the
process of low-mass star formation in the environment around massive
stars, and present a scenario for the evolution of young stars and their
disks in regions of massive star formation.  We also consider some
of the implications of the nearby presence of massive stars on the way in
which the Solar System evolved.  Our hope is to formulate this discussion
in a way that will help bridge the gap between the star formation and
the meteoritic communities, emphasizing to each that they
cannot answer the fundamental questions in their own discipline without
first appreciating the perspectives of the other.

\section{Environments for Low-Mass Star Formation}

Studies of low-mass star formation have traditionally focused on regions
such as the Taurus-Auriga molecular cloud, where low-mass Sun-like stars
form in relative isolation from each other.  There are good reasons for
this.  Located at a distance of only 140 pc, corresponding to a spatial
scale of 140 AU/arcsecond, Taurus-Auriga offers our best opportunity
to examine the low-mass star forming environment up close.  Formation
of low-mass stars in isolation is also a good place to start from a
theoretical standpoint \citep[e.g.,][]{sal87}, providing the context
for various ideas such as ``X-winds'' \citep{shu01}.  The convenience of
Taurus-Auriga does not, however, guarantee that the mode of star formation
seen there is characteristic of the way in which most low-mass stars form,
or that our Sun formed in such an environment.

\subsection{Most Sun-like Stars Form Near Massive Stars}

Several lines of evidence suggest that the majority of low-mass Sun-like
stars form in rich clusters and in relatively close proximity to
massive stars.  The mass spectrum of molecular clouds is $dN/dM \propto
M^{-\alpha}$, where measured values of $\alpha$ typically range from
1.6 to 1.8 \citep[e.g.,][]{ef96}.  This distribution is weighted toward
the most massive molecular clouds, which tend to form rich clusters
containing massive stars.  The initial mass function (IMF -- the fraction
of stars formed as a function of stellar mass) of stars in Taurus-Auriga
is also known to be peculiar in several ways \citep[e.g.,][]{gww04}.
The IMF in Taurus peaks at a higher mass ($\sim 0.8$ M$_{\odot}$) than
is common in the field, but is deficient in stars with masses greater
than about 1 M$_{\odot}$.  Taurus-Auriga also produces few brown dwarfs.
Similarly, Taurus-Auriga produces significantly more binary stars than are
seen elsewhere.  Generally speaking, field stars more closely resemble
populations that formed in rich clusters than the population forming
in Taurus.

The most direct way to address the question of where the majority
of low-mass stars form is to take an inventory of star formation in
the local neighborhood.  In their recent review, \citet{ll03} present
a survey of star formation out to a distance of 2 kpc.  They conclude
that 70-90\% of the young stars in the vicinity of the Sun formed in rich
embedded clusters.  Of the stars in embedded clusters surveyed by Lada \&
Lada, $\sim$75\% are in clusters that {\it currently} contain massive
stars, where ``massive stars'' refers to stars with masses in excess
of 8 M$_{\odot}$ that will end their lives in supernova explosions.
Other clusters may have contained massive stars in the past.

On balance, we suggest that rich clusters containing massive stars may be
a more common environment for low-mass star formation than are regions of
isolated low-mass star formation such as Taurus-Auriga.  This conclusion
differs from some previous analyses of this question.  For example,
\citep{am01} reached the opposite conclusion that more isolated, smaller
groups dominate low-mass star formation.  \citet{ll03} address this
difference in the two studies directly.  They point out that Adams \&
Myers underestimate the cluster birth rate in the Galaxy by almost an
order of magnitude.  The difficulty in the Adams \& Myers calculation
is that they fail to count clusters that disperse quickly.  Lada \&
Lada find that less than 10\% of clusters survive to an age of $10^7$
years, and that less than 4\% of clusters survive to an age of $10^8$
years.  Clusters are typically short-lived because when massive stars
disperse much of the gas in an originally bound molecular cloud, the
remains are typically unbound (see Reipurth, this volume, and Bally, Moeckel,
\& Throop, this volume).
Most clusters therefore are not included in the catalog of classical
open clusters \citep{bcd91} that \citet{am01} use to estimate the
cluster birth rate.  Quoting from \citet{ll03}, ``[This difference]
represents an enormous discrepancy and is of fundamental significance
for our understanding of cluster formation and evolution.''

\subsection{The Sun Formed Near a Massive Star}

Most stars may form in rich clusters near massive stars, but that does not
necessarily mean that the Sun formed in such an environment.  \citet{al01}
address the specific question of the likely size of the cluster in which
the Sun originated, concluding that the probability that the Sun formed
in a rich cluster is $\sim 0.85\%$.  The Adams \& Laughlin analysis,
however, is subject to the same criticism as the analysis of \citet{am01},
in that both use the same underestimate of the birth rate of clusters.
Adams \& Laughlin also argue against formation of the Sun in a cluster
environment based on the likelihood that the Solar System would have been
disrupted by close encounters with other cluster members.  In making this
assessment, however, \citet{al01} assume that clusters live for 100 times
the dynamical time scale of the cluster, or $\sim 10^8 - 10^9$ years.
This assumption leads to a substantial overestimate of the probability
of disruption if, as found by \citet{ll03}, few clusters survive even to
ages of $10^7$ or $10^8$ years.  Potential disruption of our planetary
system does not appear to place a strong constraint on whether the Sun
formed in a rich cluster.

The strongest evidence that the Sun formed near a massive star comes
from analysis of the decay products of SLRs in meteorites.  Studies of
SLRs in meteorites date back to the discovery of evidence that live
\Al\ was present in the young Solar System \citep{lpw76}.  A variety
of ideas have been put forward for the origin of SLRs in meteorites.
These include the idea, going back to \citet{ct77}, that formation of
the Sun might have been triggered by a nearby supernova which also mixed
newly synthesized isotopes into the material from which the Sun formed.
This idea has recently been considered in detail by \citet{gv00}.

Another common model for the origin of SLRs in meteorites is {\it in situ}
production in the young Solar System by spallation reactions involving
Solar cosmic rays \citep{hd76,cdw77,dwe78,lee78}.  In particular, Shu
and collaborators develop this idea within the context of their X-wind
model for low-mass star formation \citep{gou01}.  A spallation model for the formation
of SLRs has two distinct advantages.  First, it depends on no outside
source for these radioisotopes, which is a necessary condition if the Sun
formed in a region of isolated low-mass star formation similar to Taurus-Auriga.
Second, a spallation model can account for the presence of live \Be\ in
the early Solar System \citep{mcr00}.  This isotope cannot be produced by
supernova nucleosynthesis, but is produced in abundance by spallation.
This has been called a ``smoking gun'' favoring an X-wind origin for
SLRs \citep{gou01}.

The strength of arguments favoring spallation as the source of SLRs
changed dramatically when \citet{th03} reported the confirmation
of live \Fe\ in the early Solar System.  \Fe, which has a half-life
of 1.5 Myr, is a neutron-rich isotope that is not formed in abundance
by spallation reactions \citep{lee98}, but is formed in core collapse
supernovae.  The presence of this neutron-rich isotope is compelling
evidence for an interstellar origin for at least some SLRs.
Other SLRs that are difficult
to produce by spallation include $^{182}$Hf and $^{107}$Pd.  Meteoritic
evidence of gas-phase $^{36}$Cl ($t_{1/2} = 0.3$ Myr) in the outer parts
of the young Solar System has also been interpreted as presenting a
serious challenge to local irradiation models of SLRs \citep{lin05}.

Various ideas have been put forward as possible interstellar sources
for \Fe\ and other SLRs.  \citet{bgw03}, for example, suggest that an
Asymptotic Giant Branch (AGB) star may have been the source.  AGB stars,
however, are old objects and are not observed to be correlated with
sites of low-mass star formation.  Even using very optimistic assumptions,
\citet{km94} compute a probability of $< 3 \times 10^{-6}$ that the
material from which the Sun formed had recently been contaminated by
mass loss from an AGB star.  Similar statistics should apply to most other
sources involving long-lived objects in the late stages of their evolution
(e.g., novae or Type Ia supernovae).

In contrast to AGB stars, formation of massive stars {\it is} correlated
both spatially and temporally with the formation of low-mass stars.
Massive stars typically have lifetimes of $\sim$ 3--30 million years.
For realistic IMFs, \citet{ss95} find that the rate of supernovae in
a cluster will be roughly constant over the 30 million years between
the the first explosion and the time that the 8 M$_{\odot}$ stars die.
Low-mass star formation in surrounding regions is common throughout the
lives of massive stars, so many low-mass stars will experience a nearby
supernova within a few million years of their formation.  Even in a
relatively meager cluster containing only 5 massive stars, the median time
between formation of a low-mass star and a local supernova is only about
3 million years.  In the richer Scorpius-Centaurus OB association, 95\% of
the low-mass stars formed within the last 8 to 12 Myr \citep{pz99,mml02}.
During that time there have been $\sim 20$ supernovae in the association
\citep{ma01}, or roughly one every 500,000 years.

Supernova models are capable of producing most of the observed SLRs in
reasonable abundance \citep{mc00,mey03}, and it is unavoidable that
many young low-mass stars experience nearby supernovae.  We conclude
that one or more core-collapse supernovae provide {\it by far} the most
astrophysically plausible source for \Fe\ (and by association most other
SLRs) in the young Solar System.  The only well-established SLR that a
nearby supernova cannot account for is \Be.  If no other source of \Be\
were known, it might require spallation in an X-wind as a significant
source of SLRs in addition to supernovae.  An alternative source for \Be\
has been proposed, however.  \citet{dcs04} found that the amount of \Be\
in the early Solar System can be accounted for quantitatively by direct
capture of Galactic cosmic rays by the protosolar cloud, and that this may
rule out significant production of \Be\ by spallation.  Evidence of \Be\
is found in all meteorites in which it has been sought, while other SLRs
are absent in a small fraction of grains, suggesting that indeed \Be\
has a unique origin from other SLRs.  With \Be\ due to captured cosmic
rays and other SLRs consistent with or requiring a supernova origin,
it is unclear that any evidence remains requiring spallation as a source
for SLRs in the early Solar System.

We conclude that \Fe\ has replaced \Be\ as a new and unambiguous ``smoking
gun'' that places the Sun's formation in a rich cluster environment near
at least one massive star.

\section{Massive Stars Dominate the Structure of Their Environments}

\begin{figure}[!ht]
\plotone{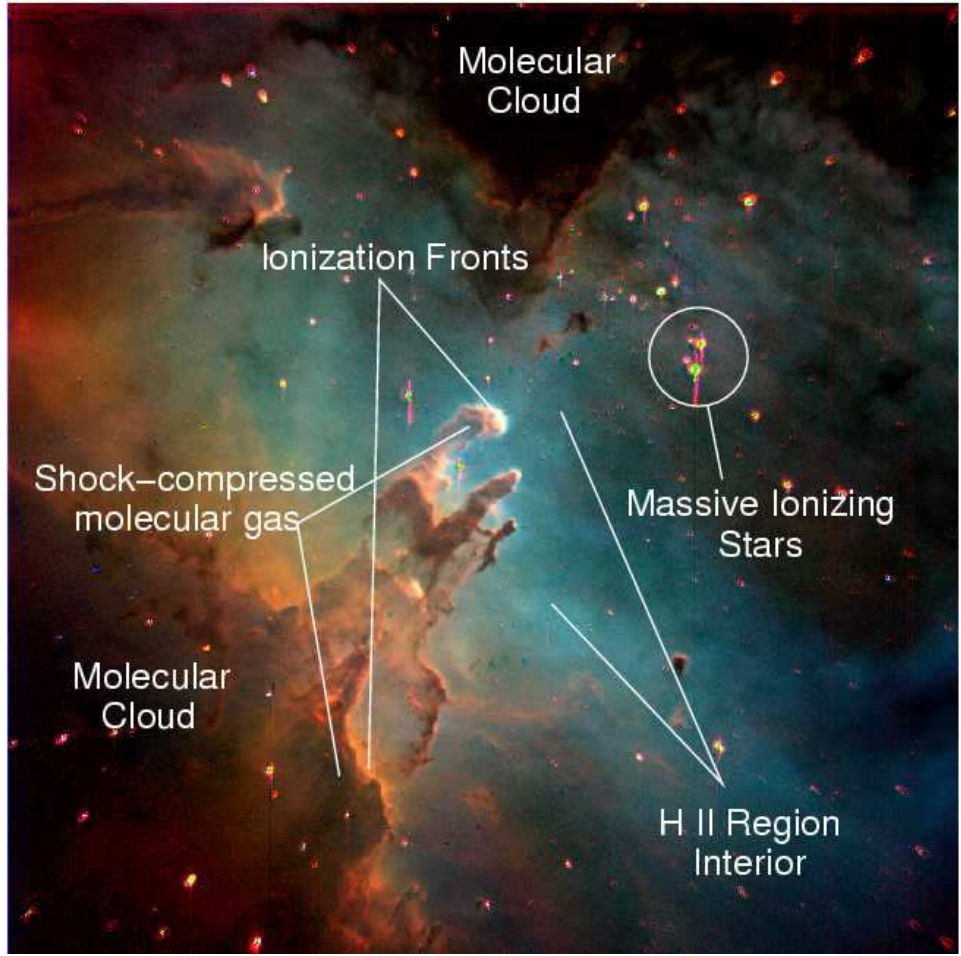}
\caption{The structure of a blister \HII\ region.  This is a ground-based
image of the Eagle Nebula, M16, obtained with the 1.5-m telescope at
Palomar Observatory.  
}
\label{EAGLE}
\end{figure}

The birth of a massive star is a violent event.  Massive stars have
extemely high luminosities, often in excess of $10^5 L_{\odot}$.
These stars produce intense extreme ultraviolet (EUV) radiation that
ionizes a star's surroundings, and intense far ultraviolet (FUV) radiation
that can penetrate into regions of neutral gas.  (The volume of ionized
gas surrounding a massive star is referred to as an ``\HII\ region,''
because the principal constituent of the region is ionized hydrogen.)
When a massive star turns on, a volume of gas around the star is rapidly
ionized and heated to $\sim 10^4$K.  The resulting high pressure region
then expands into surrounding molecular gas at roughly the speed of sound
in the hot gas.  This pressure-driven expansion typically continues
until the \HII\ region breaks free of the confining molecular cloud,
at which point the gas in the interior of the region rapidly vents into
the intercloud ISM \citep[e.g.,][]{tt82}.  The cavity that is left behind
in the wall of the molecular cloud is referred to as a ``blister \HII\
region'' \citep{zuc73}.  Figure~\ref{EAGLE} shows an image of one such
region, M16, otherwise known as the Eagle Nebula.  The interior of the
cavity of a blister \HII\ region is filled with relatively tenuous ($n \la
100$ cm$^{-3}$), hot ($T \sim 10^4$ K), ionized gas.  This gas typically
presents very little opacity to the UV radiation from the massive stars.
Of the ionizing radiation that does not escape the cavity altogether,
most reaches the wall of the cavity where it is absorbed in a very thin
zone called an ionization front.  Gas at the ionization front has yet to
expand much, so has a density like that of the molecular cloud, but has a
temperature like that of the interior of the \HII\ region.  The resulting
high pressure drives a photoevaporative flow of photoionized material away
from the surface of the molecular cloud into the \HII\ region interior.
At the same time, the pressure at the ionization front drives a shock
into the surrounding molecular cloud, compressing that gas and driving
up the pressure \citep[e.g.,][]{kah58}.  The ionization front and its
shock typically move into the molecular cloud at velocities of order
a few km/sec.  Ionization fronts are apparent in Figure~\ref{EAGLE}
as the bright rims seen at the edge of the \HII\ region.

Shocks driven in advance of ionization fronts compress dense molecular gas
surrounding \HII\ regions.  Regions in Figure~\ref{EAGLE} that have been
compressed by such shocks include several prominent columns, as well as
the dense layer of material seen immediately beyond the ionization fronts
around much of the periphery of the nebula.  Massive stars are also the
source of powerful winds, and at the ends of their lives undergo supernova
explosions that inject significant energy and momentum in addition to
newly synthesized nuclei into their surroundings.  Winds and supernovae
contribute additional momentum and energy into their environment, adding
to the compression of nearby dense molecular gas.  Once a single massive
star forms, the combined energy and momentum input from that star quickly
reshapes its environment, dominating all that goes on in its surroundings,
including the process of low-mass star formation.
Low-mass star formation around massive stars cannot be assumed to
be the mode of star formation seen in Taurus-Auriga, only writ large.
As illustrated in Figure~\ref{YSODISKS}, YSOs in \HII\ region
environments demonstrably are {\it not} ``just like'' YSOs in Taurus-Auriga.

\section{An Evolutionary Scenario for Low-mass Stars in \HII\ Regions}

\begin{figure}[!ht]
\plotone{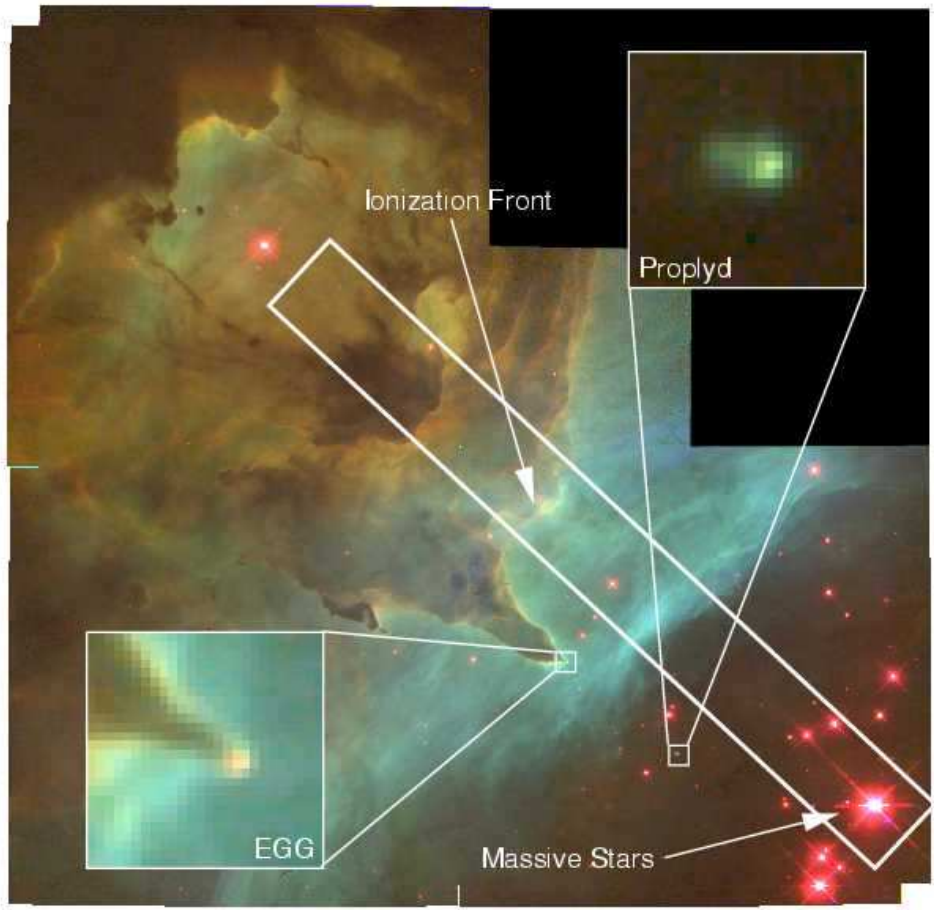}
\caption{An {\it HST} WFPC2 image of the G353.2+0.9 \HII\ region
in NGC~6357 \citep{hhss04}.  This figure illustrates the
astrophysical context for the sequence of events described in Figure~\ref{SCENARIO}.
}
\label{NGC6357}
\end{figure}

\begin{figure}[!ht]
\plotone{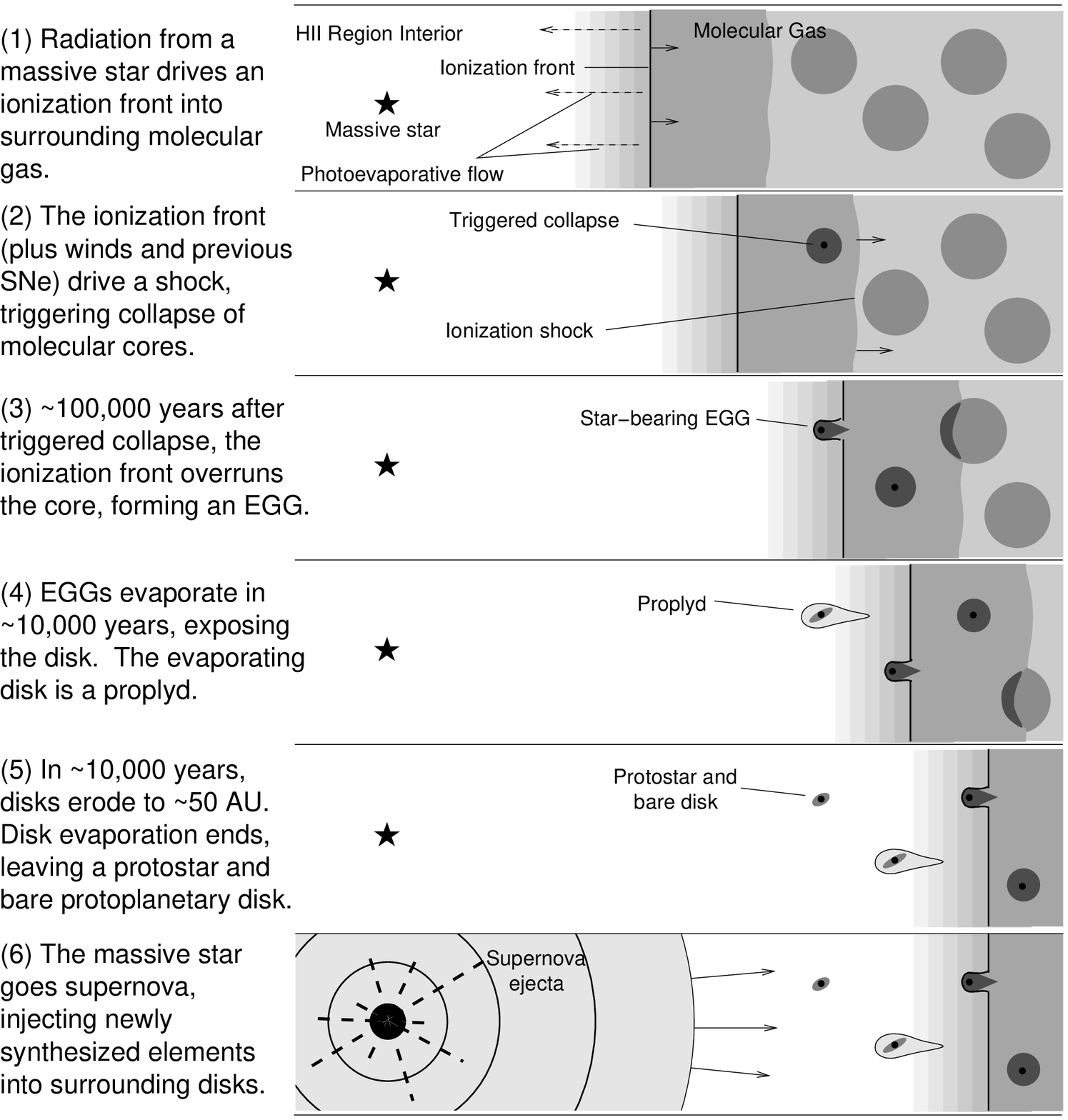}
\vskip -1in
\caption{The proposed sequence of events that characterizes the formation and
evolution of low-mass stars around the periphery of \HII\ regions.
Each panel represents a section through the nebula, as indicated in
Figure~\ref{NGC6357}.
}
\label{SCENARIO}
\end{figure}

Stars form in dense cores buried in the interiors of
molecular clouds.  Stars do {\it not} form in the hot, tenuous interiors
of \HII\ regions.  {\it The young stellar objects seen in the interiors
of \HII\ regions, such as the young stars, protostars, and ``proplyds''
seen in the Orion Nebula, did not form in the environments in which
they reside today.} Rather, these stars {\it necessarily} formed in
the dense molecular gas that once surrounded the massive stars and were
later uncovered by the advancing ionization front.  As a young star and
its natal environment are overrun by an ionization front and emerge
into the interior of an \HII\ region they go through a well-ordered
evolutionary sequence.  This evolutionary sequence is illustrated
in Figures~\ref{NGC6357} and \ref{SCENARIO}.  Figure~\ref{NGC6357}
shows an {\it HST} WFPC2 image of the G353.2+0.9 \HII\ region in
NGC~6357 \citep{hhss04} with a strip indicated cutting from the massive
stars across the ionization front and into the molecular gas beyond.
Figure~\ref{SCENARIO} shows schematically the sequence of events that
are experienced by young stellar objects as the ionization front pushes
into the molecular cloud.  We suggest that this sequence provides the
correct astrophysical context for understanding the early evolution of
the Solar System.

\subsection{Gas is Compressed Around \HII\ Region Peripheries}

In Figure~\ref{SCENARIO} panel 1, radiation from a massive star drives
an ionization front and its leading shock into surrounding molecular gas.  
(Momentum input from winds
and supernovae contribute to the momentum input into the cloud as well,
but this does not change the basic picture of energy input from a massive
star compressing the molecular gas beyond the edge of the \HII\ region.  Nor
does it change the basic nature of the ionization front that defines that
interface.)
Compression around the periphery of \HII\ regions is observed in
many ways.  Observations of the molecular gas itself often shows 
compression just outside the ionization front, together with the kinetic signature
of expansion \citep[e.g.][]{whi99}.  Observations of photodissociation
regions (regions immediately beyond the ionization front where H$_2$ is
dissociated by FUV radiation) show these regions to be at significantly
higher pressure and the ambient molecular cloud \citep[][and references
therein]{ht97}.  Observations of magnetic fields around \HII\ regions
show that the fields generally line up with the edge of an \HII\ region,
which is a clear signature of compression perpendicular to the field
\citep[e.g.,][]{gae01}.  Compressed molecular gas around \HII\ regions can
often also be seen in extinction against the background of the nebula,
as described above.

\subsection{Star Formation Can be Triggered by Compression}


It is reasonable that significant compression of the gas around
the periphery of an \HII\ region might play a role in initiating
the collapse of clumps and the formation of stars in this region
\citep[e.g.,][]{el77,ber89,bm90}, as shown schematically in Figure
\ref{SCENARIO}, panel 2.  It is possible that, left to themselves, some of
these clumps eventually might have collapsed on their own.  These clumps
are not left to themselves, however.  They are located in gas that will
soon be dispersed by the advancing ionization front.  Without the added
push from the high post-shock pressure, it is likely that these clumps
would have been destroyed before collapsing to form stars.  The clump
mass required for collapse also decreases with increasing pressure.
The mass required for the collapse of a virialized sphere, for example,
is $\propto P^{-1/2}$, where $P$ is the pressure of the surrounding
medium \citep[e.g.,][]{lar03}.  There should therefore be a population
of low-mass clumps that would not have collapsed at all were it not for
compression in advance of the ionization front.

Numerous authors have argued in favor of triggered star formation
around \HII\ regions on the basis of protostars located in gas that has
been compressed by the expansion of the \HII\ region.  For example,
\citet{lc00} find that star formation around the periphery of M20 is
concentrated in cores undergoing radiatively driven implosion, while star
formation in W5 lies just beyond the boundaries of the \HII\ region in
gas that shows the kinematic signature of ionization shocks \citep{km03}.
\citet{hhc04} recently found that water masers in M16, which trace the
presence of Class 0 protostars, are concentrated in compressed molecular
gas located just beyond ionization fronts.  
Similarly, \citet{rea04} found that protostars in the cometary globule
IC~1396 seen in {\it Spitzer} IRAC and MIPS 24 $\mu$m images are
concentrated within 0.02 pc of the ionization front on the leading edge of
the globule.  \citet{pz99} argue that low-mass star formation throughout
the Upper Scorpius OB association was triggered.  While \citet{dm01}
stop short of claiming that star formation is triggered, the sequence of
ages and properties of stellar populations they report in their study of
$\lambda~Orionis$ is in very good agreement with what one would expect
if star formation throughout the region was triggered by the expansion
of the \HII\ region.  The list of specific regions in which various
authors have argued for triggering of low-mass star formation by massive
stars could go on at length.  This issue is discussed further below in
Section \ref{TESTABLE}

In this paper we stress the role of massive stars in triggering
{\it low-mass} star formation, as is appropriate to the focus of
this volume.  However, energy input from massive stars plays a key
role in triggering the formation of additional {\it massive} stars as
well \citep[e.g.,][]{bla91}.  \citet{oey05}, for example, find that two
subsequent generations of massive stars in W3/W4 were triggered by an
original generation.  Each of these generations of massive stars in turn
gave rise to large populations of triggered low-mass star formation.

\begin{figure}[!ht]
\plotone{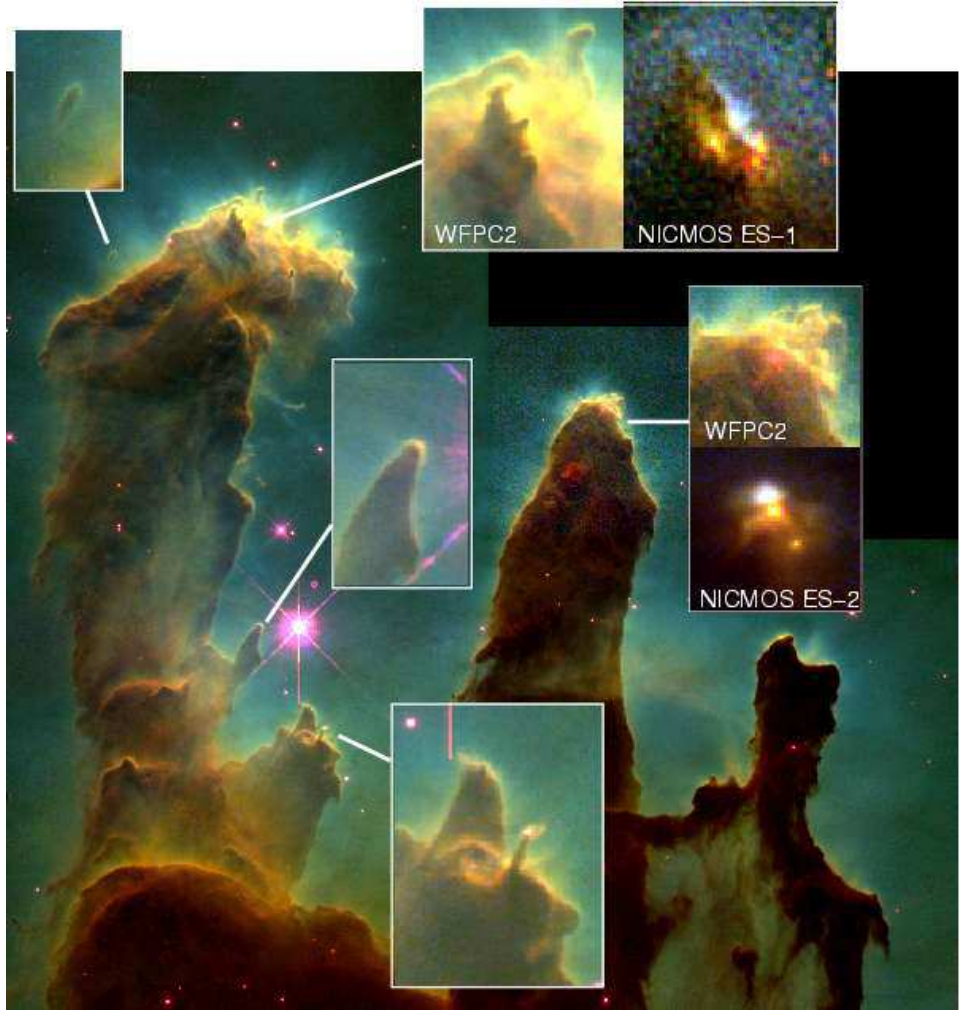}
\caption{{\it HST} WFPC2 image of the three central columns in M16
\citep{hes96}, along with enlargements of several EGGs and {\it HST}
NICMOS images of two prominent YSOs \citep{tsh02}.}
\label{EAGLEHST}
\end{figure}

\subsection{Emerging Clumps are Seen as EGGs}

The next stage in the evolution of a young star's environment comes
when the ionization front overruns a clump and the young star within,
as illustrated in Figure~\ref{SCENARIO}, panel 3.  The ionization front
typically moves into the cloud at $\sim$ 1-2 km/sec, while the typical
separation between the shock and the ionization front is $\sim$ 0.1 pc,
so the ionization front will overtake young stars and dense clumps within
$\sim 10^5$ years after passage of the shock.  As
the ionization front sweeps past a dense clump and the young star within
(if one has formed), the evaporating clump appears as a small protrusion
extending from the molecular cloud into the interior of the \HII\ region.
The clump may even become detached altogether from the wall of the
molecular cloud.  This is the ``EGG'' or ``evaporating gaseous globule''
phase of evolution of the object.  The EGG phase of evolution was first
seen in {\it HST} WFPC2 images of M16 \citep{hes96}.  Several EGGs in this
field have visible stars at their tips.  Figure~\ref{EAGLEHST} shows that
image together with enlargements of a number of EGGs that are known to have
associated stars.  Also included are {\it HST} NICMOS near infrared images
of young stellar objects at the heads of Columns 1 and 2 \citep{tsh02}.

In the previous section we argue that concentration of young stars
in gas that has been compressed by passage of the ionization front is
evidence of triggering of star formation by an expanding \HII\ region.
The $\sim 10^5$ years between passage of the shock and passage of the
ionization front is comparable to the timescale over which low-mass
stars accrete their mass.  This means that arrival of the ionization
front and photoevaporation of a protostar's surroundings may cut the
protostar off from its accretion reservoir prematurely, and may be an
important factor in limiting the final mass of the star.  This idea
was discussed in the context of {\it HST} images of M16
\citep{hes96}, and was later proposed independently in the context of
the Orion Nebula \citep{rob04}.  Among the predictions made on the basis
of this idea is that the number of brown dwarfs would be found to be
higher in regions of massive star formation than in regions of isolated
star formation \citep{hes97,wz04}.  Observations subsequently confirmed this
prediction \citep[e.g.,][]{gww04}.  Limiting the masses of protostars by
photoevaporation of accretion reservoirs might also push the peak of the
IMF toward lower mass, which is qualitatively in the right direction to
explain the observed differences between the IMF in Taurus and the IMF
in field stars and in rich clusters.

Observations of accreting protostars that are about to be overrun by
ionization fronts support a role for photoevaporation in limiting the
masses of stars.  Examples include the Class 0 protostars associated with
water masers discussed by \citep{hhc04},
and the very young
YSOs in IC~1396 observed by \citet{rea04}.  In each case, very young
objects likely to still be experiencing infall are seen in regions that
will be overrun by ionization fronts within a few $\times 10^4$ years.
The statistics of EGGs are consistent with truncation of infall by
photoevaporation as well.  If all or almost all EGGs contained visible
stars, it might indicate that the ionization front was uncovering a region
in which star formation was a {\it fait accompli}.  \citet{ma02} find
instead that $\sim 15\%$ of EGGs in M16 contain young stars that can be
seen at 2 $\mu$m from the ground, as might be expected if the ionization
front is overrunning a region in which star formation is still ongoing.

\begin{figure}[!ht]
\plotone{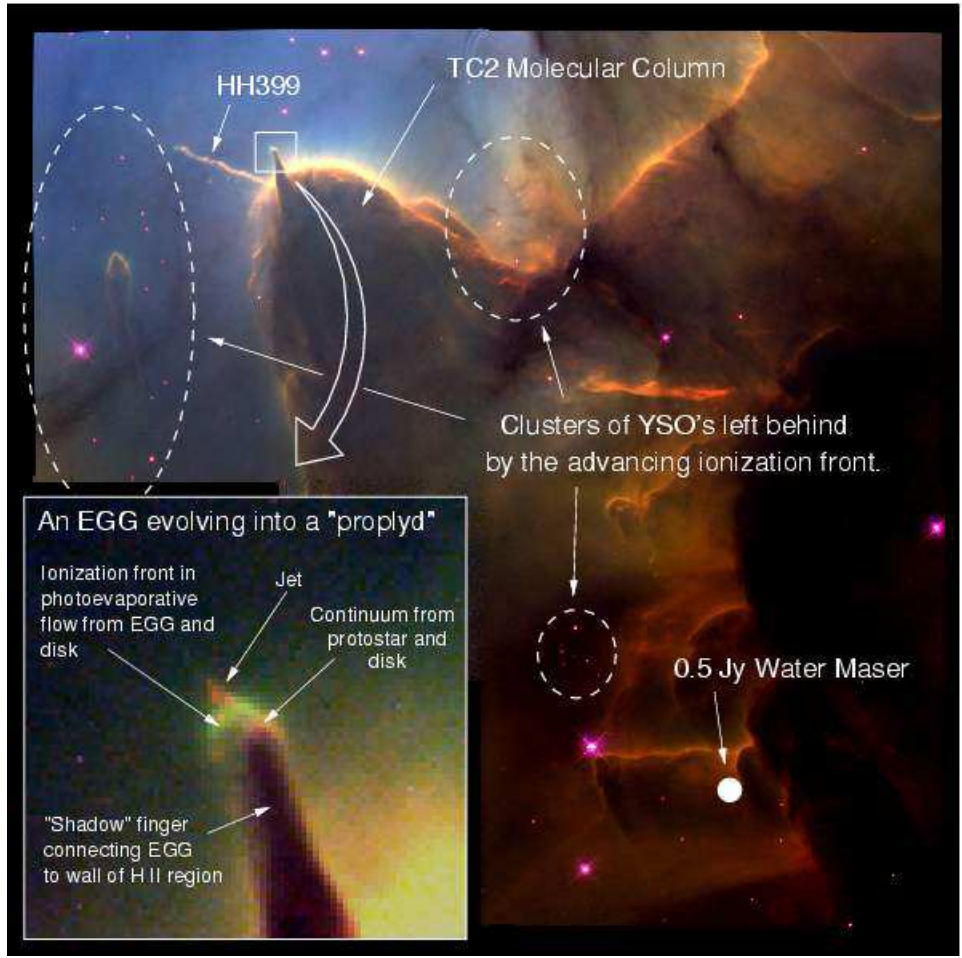}
\caption{{\it Hubble Space Telescope} WFPC2 image of a field in the Trifid
Nebula \citep{hes04}.
}
\label{TRIFID}
\end{figure}

\subsection{Photoevaporating Disks are Seen as Proplyds}

An EGG is photoevaporated within $\sim 10^4$ years of the
passage of the ionization front.  If there is a young star inside the
EGG, the star and its disk are then exposed to the intense UV radiation
from the nearby massive stars, as shown in Figure~\ref{SCENARIO},
panel 4.  Gas that is photoevaporated from the surface of the disk by
FUV radiation is ionized by EUV radiation a short distance above the
disk, and pushed back into a characteristic teardrop shape 
\citep[e.g.,][]{jhb98}.  The EGG has become a ``proplyd.''  

The best-known examples of the evaporating disk or proplyd phase of
evolution are seen in the Orion Nebula 
(see Figure~\ref{YSODISKS}b).  Orion provides the
best opportunity to see these objects, simply because is the nearest
prominent \HII\ region.  Proplyds have now been seen in a number of
other \HII\ regions, including the Carina Nebula \citep{sbm03}, and
NGC~6357 (Figure~\ref{NGC6357}).  Not all proplyds are easily resolved, even with
{\it HST}, but H$\alpha$ excesses in stellar spectra can indicate the
presence of unresolved photoionized gas associated with evaporating disks.
Studies of the properties of young stars in \HII\ region environments
\citep[e.g.,][]{dm01,sug02} typically find that young stellar objects
with H$\alpha$ excesses are most common near ionization fronts,
while weak line YSOs are located deeper in the interiors of \HII\ regions.
This trend is at least in part due to destruction of circumstellar disks.

We propose in our scenario that EGGs and proplyds are two evolutionary
phases objects pass through as they are uncovered and left behind by the
passage of ionization fronts.  If this is the case then we would expect
to see objects that are undergoing the transition from EGG to proplyd.
{\it HST} images show a number of such objects.  A good example can be
seen in the inset in Figure~\ref{TRIFID} \citep{hes04}.  The tip of this
finger-like EGG in the Trifid Nebula shows a number of features that are
characteristic of proplyds.  These include reflection nebulosity from
the star at the tip of the EGG, an ionized evaporative flow standing off
from the star by $\sim 10^{16}$ cm, and a small stellar jet.  The lower
part of this object is a classical EGG of the sort seen in M16, but the
upper portion of this object has all of the characteristics of an Orion
proplyd \citep{bom00}.  A similar object can be seen in the original {\it
HST} image of M16 as well, where a small [S~II] jet protrudes from the
region around a visible star sitting at the tip of a long finger-like EGG
(bottom inset in Figure~\ref{EAGLEHST}).

\subsection{Proplyds Leave Behind Young Stars with Truncated Disks}

Proplyds are also short-lived objects.  Within $10^4$ -- $10^5$ years
disks are eroded down to a radius of 30 AU or so.  Photoevaporation of the
rest of the disk takes several million years \citep{jhb98,sh99}.
As shown in Figure~\ref{SCENARIO}, panel 5, after the outer parts of
the disk are evaporated, the protostar and its now-truncated disk are left
sitting in the hot, tenuous interior of the \HII\ region, where the disk
remains subject to intense ultraviolet radiation.

This is the longest-lived phase in our scenario, and should therefore be
the most common stage in which to find young stellar objects in
\HII\ region environments.  Observations bear this out.  For example,
there are around 4,000 young low-mass stars in the volume of the \HII\
region ionized by the Trapezium cluster in Orion \citep{hh98}.  Of these,
\citet{hil98} find that 55--90\% have infrared excesses indicating
that they are still surrounded by disks.  In contrast, only about 150
objects are seen in Orion in the EGG or proplyd phases \citep{ode98}.
This is easily understood if truncated disks live for $\ga 20$ times as
long as do proplyds, in reasonable agreement with the lifetimes
estimated above.  These statistics should be viewed with caution because of
the different selection criteria used in different studies.  Even so, they
make the point that as expected, proplyds are relatively rare as compared
with low-mass YSOs surrounded by truncated disks.

In our scenario, progress of the ionization front should leave behind
small groups of young stars that formed in large cores that were
compressed by radiatively driven implosion and then disrupted by
photoevaporation.
A number of such small groups of stars can be seen in the {\it HST}
image of the Trifid Nebula in Figure~\ref{TRIFID}.  Of particular
interest is the group of stars on the left side of the image that
surround the tadpole-shaped remains of the molecular column from which
the stars formed.  There are many indicators of ongoing star formation in
the TC2 molecular column in the central part of the image \citep{lef02},
including a prominent stellar jet (HH 399) emerging from an unseen young
stellar object located near the head of TC2, and a water maser indicating
the presence of a young protostar \citep{hhc04}.  It seems likely that in
$10^5$ years or so TC2 will look much like the tadpole and its surrounding
group of young stars on the left edge of the field of view.

Figure~\ref{SCENARIO} panel 5 shows the environment where a Sun-like
star and its protoplanetary disk spend most of their youth, immersed
in the hot, low-density interior of an \HII\ region, subject to intense
ultraviolet radiation and possibly fast winds and other effects due to
nearby massive stars.  {\it This is
where formation of our own Solar System is likely to have begun.}

\subsection{Protoplanetary Disks May Be Overrun by Supernova Ejecta}

\begin{figure}[!ht]
\plotone{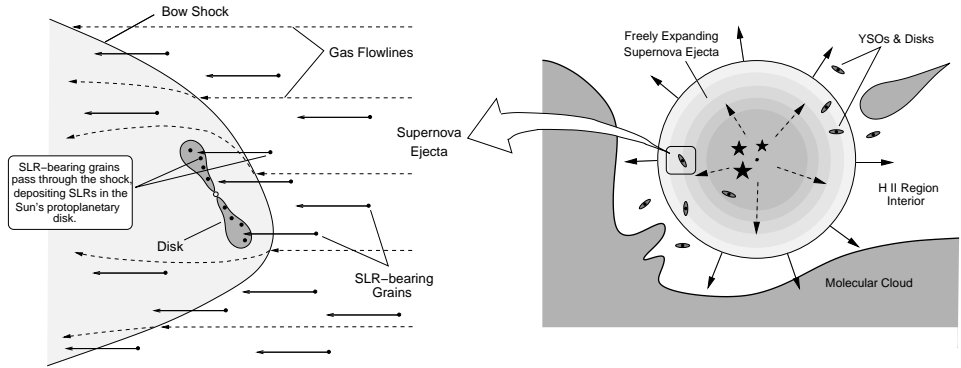}
\caption{Illustration of the ``aerogel'' model for injection of SLRs
from Ouellette, Desch, \& Hester (this volume).}
\label{AEROGEL}
\end{figure}

A final entry in the sequence of events befalling a young low-mass star in
an \HII\ region environment is shown in Figure~\ref{SCENARIO}, panel 6.
Massive stars live short lives, typically $\sim 3-30$ million years,
then die in supernova explosions.  When a supernova goes off in an \HII\
region the ejecta will expand freely through the low-density cavity,
sweeping over any low-mass stars and their disks present at that time.
When the gaseous ejecta from a supernova hits a protoplanetary disk,
a bow shock is established that directs most of the ejecta around
the disk.  \citet{chev00} discussed this possibility, and showed that
a protoplanetary disk around a young star should survive passage of the
supernova ejecta relatively unscathed.  Some turbulent mixing may occur
as the ejecta sweeps past the disk, and some capture of ejecta by the
gravitational potential of the young star is possible.  However, even
if very little of the gaseous supernova ejecta winds up in the disk,
dust in the ejecta will continue to travel on ballistic trajectories,
passing through the bow shock then being vaporized as it encounters the
much greater column of gas in the disk itself.  This ``aerogel'' model
for the injection of short lived radionuclides into protoplanetary disks,
illustrated in Figure~\ref{AEROGEL}, is discussed by Ouellette, Desch,
\& Hester (this volume).
Dust is known to form in supernova remnants.  Its presence is inferred
from infrared emission from supernovae themselves \citep[e.g.,][and
subsequent work on SN1987A]{luc89}, and can even be seen directly in
absorption against the background of the synchrotron emission in the
Crab Nebula \citep[e.g.,][]{fb90,san98}.  Many of the SLRs found in
meteorites are refractory elements that might be expected to be found
in the solid phase in supernova ejecta.

The interaction of supernova ejecta with a low-mass star and
protoplanetary disk has not been observed directly, which is not
surprising as these are fleeting events.  Even so, the occurrence of such
events should not be considered controversial.  Most low-mass stars seen
in \HII\ regions are found in 
close proximity to stars that {\it will} go supernovae in the relatively
near future, with very little intervening material to stand in the way
of the expanding ejecta.  As Ouellette, Desch,
\& Hester (this volume) show, this provides a natural
way to inject SLRs into the protosolar disk in reasonable abundances to
explain the meteorite data.

There are several caveats worth mentioning at this point.  The first is
that most clusters disburse on time scales that are comparable to the
lifetimes of massive stars \citep{ll03}.  While we have not addressed the
effects of migration of either the high- or low-mass stars quantitatively,
this will not change the basic conclusion that many low-mass stars will
be close proximity to a supernova explosion within an a few million
years of their formation.  {\it Proximity to a supernova explosion is
an unavoidably common aspect of the evolution of low-mass stars.}  Even so,
there is need for much more detailed work on cluster dynamics and related
topics if we are to understand the overall statistics of SLR injection
into protoplanetary disks.

A second caveat is that there may be alternative paths by which SLRs can
find their way from massive stars into protoplanetary disks in and around \HII\ regions.
Supernova injection of SLR-bearing dust into the molecular cloud is
likely when ejecta hits the wall of a cavity.  This newly synthesized
material could find its way into forming planetary systems located very
near to the ionization front.  These are much less common, however, than
young stars and disks located in the interiors of \HII\ regions.  The
Wolf-Rayet phase of massive star evolution is also characterized by significant
mass loss, and may provide another source for newly synthesized material
if a way can be found to mix that material into protoplanetary disks.
Neither of these issues change the idea underlying our basic scenario
for late injection of SLRs, but rather allow for the possibility of
additional avenues whereby material from an evolved massive star might
find its way into a protoplanetary disk.

\section{A Closer Look at Triggered Star Formation}

The scenario above outlines the sequence of events that takes place
as an ionization front overruns a region of low-mass star formation.
For the most part, this sequence of events does not depend on whether
low-mass star formation is triggered or not.  {\it Any young object that finds
itself in the interior of an \HII\ region must have experienced the basic
sequence of events illustrated in Figure~\ref{SCENARIO}.}  While perhaps
of secondary importance to the meteorite community, however, the question
of how low-mass star formation is initiated is of vital importance to
astrophysics.  

As discussed by Bally, Moeckel,
\& Throop (this volume) and Reipurth (this volume), there has been much
recent progress on the process of star formation in a hierarchy of
nested density fluctuations resulting from turbulence in molecular
clouds \citep[][and references therein]{vaz04}.  We do not rule out the
possibility of hierarchical collapse in turbulent gas that has been
compressed around \HII\ regions.  Nor do we rule out the possibility
that ongoing hierarchical collapse might have shaped the spectrum of
clumps that are overrun by ionization fronts and their leading shocks.
From our perspective the crucial issue is not the detailed mechanism
by which compression triggers star formation around \HII\ regions, but
rather the timing of that star formation and the likelihood that forming
stars will soon be overrun by an ionization front.
The key question, then, is how to
assess the relative importance of triggered star formation as opposed to
star formation that occurs near to but independently of massive stars,
and is subsequently uncovered by the passage of an ionization front.

\subsection{Triggering of Star Formation is a Testable Hypothesis}
\label{TESTABLE}

\begin{figure}[!ht]
\plotone{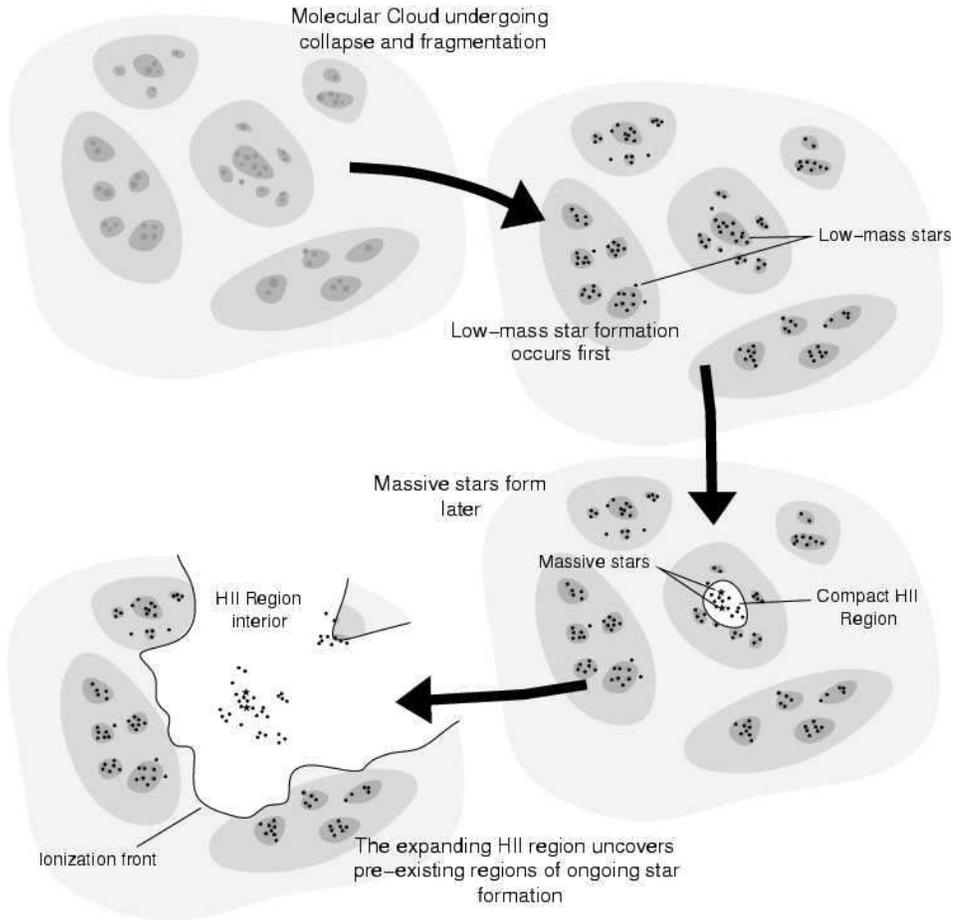}
\caption{The evolution of an \HII\ region environment, assuming that most
low-mass star formation is not influenced by the
massive stars. Compare with Figure~\ref{TRIGGERED}.
}
\label{UNTRIGGERED}
\end{figure}

\begin{figure}[!ht]
\plotone{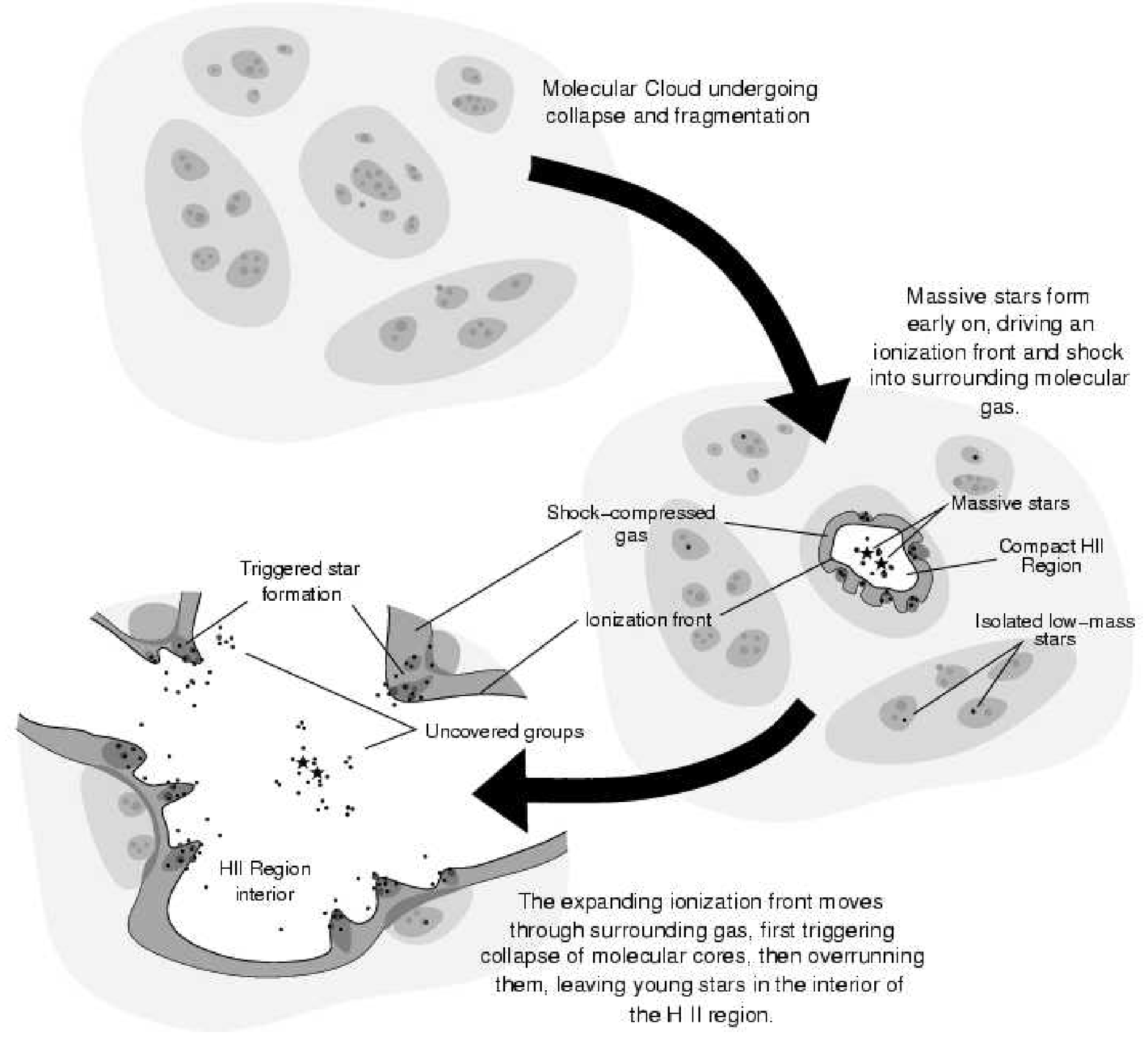}
\caption{The evolution of an \HII\ region environment, assuming that most
low-mass star formation in the region is triggered by expansion of the \HII\
region.  Compare with Figure~\ref{UNTRIGGERED}.
}
\label{TRIGGERED}
\end{figure}

Evidence cited for triggered star formation in specific regions often
amounts to finding so many young stars in compressed gas around an
\HII\ region that their presence there cannot be attributed to chance.
If applied systematically, this same criterion could be used to
investigate the extent of triggered low-mass star formation around \HII\
regions as compared with star formation in clouds that collapse in the
absence of external triggers.  If most low-mass star formation takes
place independently of the effects of massive stars, then low-mass star
formation should be uncorrelated with compressed gas around \HII\ regions.
Instead, rich regions of low-mass star formation should then be seen in
the extended regions around \HII\ regions, especially young and compact
\HII\ regions, in gas that is unaffected by \HII\ region expansion.
In this case, which is illustrated in Figure~\ref{UNTRIGGERED}, we also
expect to find regions of low-mass star formation that are comparably
intense to those found in \HII\ region environments, but in regions that
have yet to form {\it any} massive stars.  Finally, we would expect that the
ages of many low-mass stars found in \HII\ regions would be greater than
the ages of the massive ionizing stars.
Conversely, if most low-mass star formation in regions around massive
stars {\it is} triggered, then young stellar objects should be found
concentrated in gas that has been compressed by \HII\ region expansion,
as illustrated in Figure~\ref{TRIGGERED}.  We would also expect low-mass
stars in \HII\ regions to typically have a spread of ages up to the age
of the massive stars.

It is clear that some star formation is triggered by massive stars,
while other star formation is not.  Figures~\ref{UNTRIGGERED} and
\ref{TRIGGERED} suggest a means by which the relative contributions
of these two modes of star formation might be assessed.  A number of
new data sets such as the {\it Spitzer} GLIMPSE survey of the inner
portion of the Galactic plane may allow this test to be carried out for
a statistically meaningful number of objects.

\begin{figure}[!ht]
\plotone{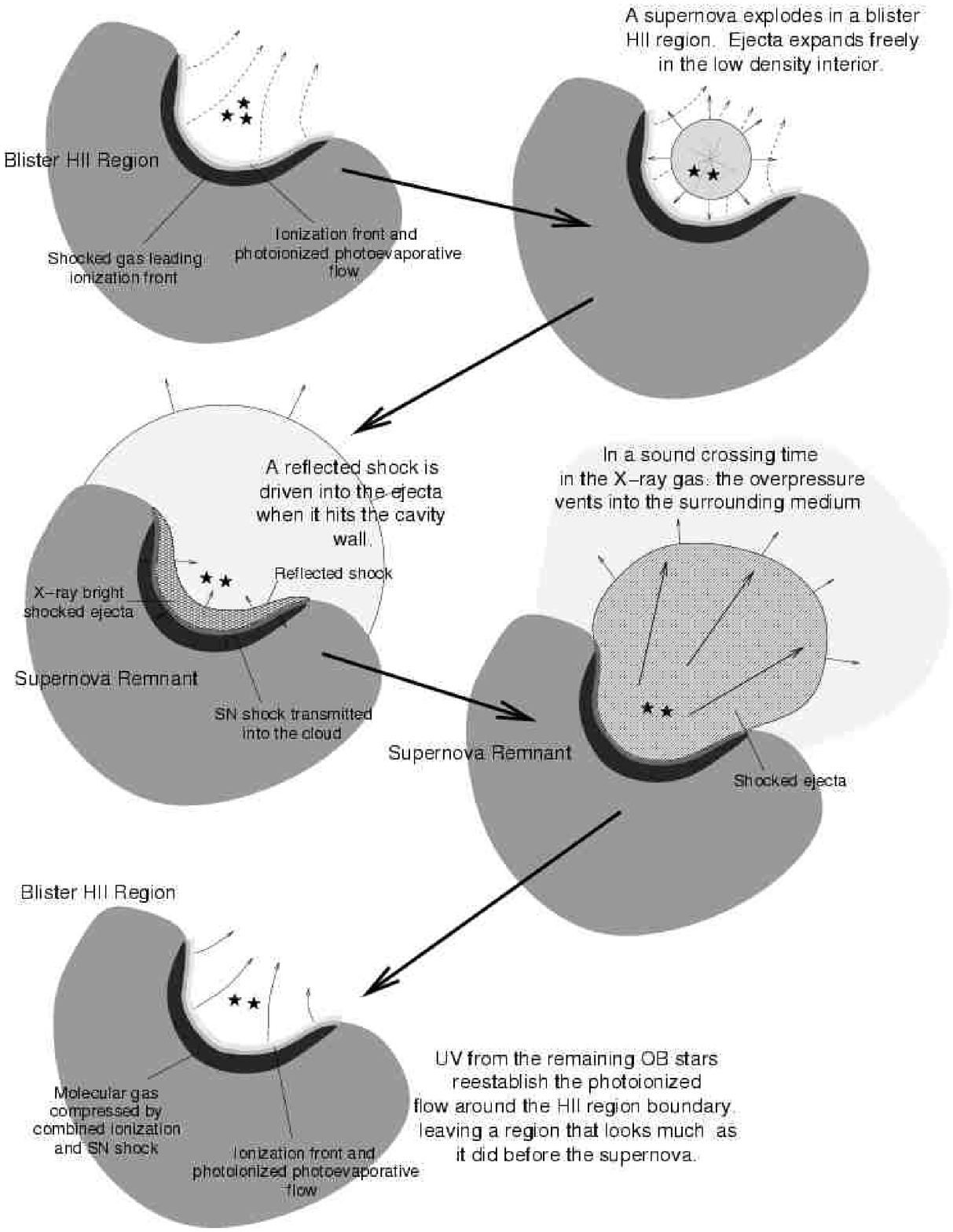}
\caption{The evolution of an \HII\ region environment following the
explosion of one of the massive stars in a supernova.}
\label{SNR}
\end{figure}

\subsection{This is Not Classical ``Supernova Triggered Star Formation''}

The mode of induced star formation discussed here differs from
``classical'' supernova triggered star formation as understood by either the
star formation or meteorite communities.  To star formation researchers,
supernova triggered star formation refers to the idea that a massive
star mostly triggers star formation in its surroundings with the sudden
burst of energy of a supernova explosion.  While supernovae contribute
to the compression of gas around massive stars, they are not necessarily
the dominant means by which star formation is induced.

As discussed above, the radiation and winds from a massive star carve a
large cavity in the interstellar medium, compressing the surrounding gas
into a dense shell.  Even when a supernova does occur, the energy released
is unlikely to dominate the cumulative effects of radiation and winds.
Figure~\ref{SNR} shows what happens when a supernova explodes in a
blister \HII\ region.  (This is the type of astrophysical environment
in which many massive stars are found and in which intense low-mass
star formation is seen, so it is the appropriate environment in which to
consider the effects of supernovae.)  When a blister \HII\ region forms,
the high pressure interior breaks out of the molecular cloud and vents
into the surrounding lower density ISM, as shown in Figure~\ref{SNR}a.
When the supernova goes off, the ejecta expands freely through the low
density cavity (Figure~\ref{SNR}b).  This is the phase during which SLRs
may be injected into protoplanetary disks (Figure~\ref{AEROGEL}).  Some of
the ejecta leaves the \HII\ region directly, but some hits the wall of
the cavity.  Where this happens, a reverse shock is driven back into the
ejecta, thermalizing its kinetic energy (Figure~\ref{SNR}c).  The sound
speed in the gas between the molecular cloud and the reflected shock
is comparable to the velocity of the ejecta, or a few thousand km/sec.
This high pressure region expands, venting out of the cavity in a time
comparable to the sound crossing time for the cavity at the temperature
of the shocked ejecta (Figure~\ref{SNR}d).  For $R_{HII} \sim 5$ pc and
$c_s \sim 2000$ km/sec, this corresponds to about 5,000 years between the
time of the supernova and the time at which the pressure in the interior
of the region has dropped significantly.  Once the overpressure associated with the
supernova has vented, any remaining massive stars will reestablish a
photoevaporative flow away from the walls of the cavity, and the region
will soon look much like the \HII\ region it was prior to the explosion
of the supernova (Figure~\ref{SNR}e).

In an event like that illustrated in Figure~\ref{SNR} little of the
kinetic energy or ejecta mass from the supernova are trapped within
an \HII\ region or transfered to the surrounding molecular gas.
Many well-studied supernova remnants have been inferred to be such
``cavity explosions,'' and extended regions of hot gas are observed to be
associated with regions of massive star formation.  The Eridanus bubble,
for example, is filled with hot gas that has vented from star forming
regions in Orion.  Observations of $\gamma$-ray line emission from \Al\
in the direction of Eridanus offer direct evidence that this bubble
contains recently synthesized supernova ejecta \citep{die02}.  Similarly, it has been
proposed that the Local Bubble is filled with hot gas originating in
the Sco-Cen OB association, and that ejecta from supernovae in Sco-Cen
are responsible for \Al\ found in ocean floor sediments on Earth.

Calculations of classical supernova triggered star formation typically
assume that the pressure of the thermalized ejecta is confined within a
small volume around the site of the supernova, and that this confined high
pressure gas has a long time to do work on its surroundings. 
This assumption leads to a significant overestimate of the amount of compression
caused by the supernova.  If the explosion takes place within a blister,
as described above, the momentum deposited at the wall of the cavity is
close to that required to reverse the motion of the ejecta, or
$\Delta p_{SN} \sim 2 M_{ejecta} \langle v_{ejecta} \rangle/4 \pi R_{HII}^2$,
where $M_{ejecta}$ is the mass of the supernova ejecta, 
$\langle v_{ejecta} \rangle$ is the average speed of the ejecta, and
$R_{HII}$ is the radius of the \HII\ region.
The momentum deposited by the ionization front is given by the pressure
at the ionization front times the lifetime of the \HII\ region, or
$\Delta p_{IF} \sim 2 n_{e,IF} k T_{IF} t_{HII}$, where $n_{e,IF}$ is the
electron density in the ionization front, $T_{IF}$ is the temperature in
the ionization front, and $t_{HII}$ is the lifetime of the \HII\ region.
For characteristic values ($n_{e,IF} \sim 5,000$ cm$^{-3}$, $T_{IF} \sim
8000$ K, $M_{ejecta} \sim 10 M_{\odot}$, and $\langle v_{ejecta} \rangle
\sim 3,000$ km/sec), we find

$$
{{\Delta p_{IF}} \over {\Delta p_{SN}}} \sim 3.5 \left({ {R_{HII}} \over
{1~{\rm pc}}}\right)^2  \left({ {t_{HII}} \over {1~{\rm Myr}}}\right).
$$

For very small, short-lived \HII\ regions, a supernova may come to dominate
the momentum input due to radiation, but in such regions stellar
winds would likely dominate both.  On the other hand, for an \HII\ region
with a radius of 5 pc around a star with a lifetime of 2 Myr, the momentum
deposition in the surrounding molecular material due to radiation will be
two orders of magnitude greater than that due to a supernova explosion.
While it is true that the injection of momentum from a supernova is very
abrupt and will lead to a temporary spike in the pressure, supernovae
seem unlikely to be the principle way in which massive stars trigger
low-mass star formation in their surroundings.

``Supernova triggered star formation'' takes on a more specialized meaning
when discussed in the context of meteorites.  A possible explanation for
the injection of SLRs into the Solar disk, dating back to \citep{ct77},
is that a {\it single} supernova event both triggered the collapse of
the Sun {\it and} injected SLRs into the material from which the Sun
formed.  We stress that our proposal is fundamentally different from
this picture.  While models of this process suggest a single supernova
could both trigger collapse and seed a clump \citep[e.g.,][]{vb02}, the
astrophysical setting required (a slow shock with all of the supernova
ejecta trailing it) is unlikely to be as common as the astrophysical
setting in which supernova ejecta directly impacts protoplanetary disks
in \HII\ region interiors.

\section{Implications of the Birth Environment of the Sun}

Most low-mass stars form near massive stars.  Analysis of SLRs in
meteorites shows that the only habitable system of which we know --
ours -- was among them.  Star formation in these environments is
demonstrably {\it not} just a scaled-up version of the same processes
seen in nearby regions of isolated low-mass star formation.  Our hope
is that by providing a specific testable scenario for low-mass star
formation around massive stars, that we might encourage redirection
of some of
the research effort that has for so long been focussed on regions like
Taurus-Auriga.  At the same time, we hope to encourage the meteorite
and planetary science communities to consider the ways in which nearby
massive stars might have affected the young Solar System.

Studies of low-mass star formation in \HII\ regions hold the promise
of new insights into a number of fundamental astrophysical questions
including, for example, the origin of the IMF.  These insights may apply
on the much larger scales of the giant \HII\ regions that dominate star
formation in many other galaxies.  \citet{sco98}, for example, found
that the physical conditions (radiation field, pressure, temperature,
ionization stratification, and morphology) at ionization boundaries in
{\it 30 Doradus} are quite similar to those found in smaller, nearby
\HII\ regions.  It is reasonable to assume that the same processes
forming low-mass stars in our neighborhood are at work in regions such
as {\it 30 Doradus} as well.

The question of low-mass star formation near massive stars is of even
greater importance to the planetary science community.
EUV and FUV radiation from
nearby massive stars incident on the Sun's protoplanetary disk is likely
to dominate that from the Sun itself.  This radiation may have played an
important role in the chemistry of the disk.  For example, UV radiation
may have been a driver of the complex chemistry required to explain the
properties of organic compounds in carbonaceous chondrites \citep{ber99}.
Massive stars would also provide a convenient source of the FUV radiation
that \citet{yl03} invoke to account for observed anomalies in the relative
abundances of $^{16}$O, $^{17}$O, and $^{18}$O \citep{cla03}.

Turning to the overall structure of the Solar System, truncation of the
Sun's protoplanetary disk to a size of a few tens of AU provides a
natural explanation for the observed 50 AU outer edge of the Kuiper
Belt \citep{cb99}.  This has also been invoked as a potential explanation
for the relatively low masses of Uranus and Neptune \citep{sjh93}.
Short disk lifetimes near massive stars may provide support for
unstable collapse rather than core accretion as the origin for giant
planets \citep[e.g.,][]{boss01,boss03}.  Dynamical effects of a cluster
environment on the young Solar System might also have been important.
The orbit of Sedna, for example, is most easily understood if it was
disturbed by an encounter with another cluster member \citep{btr04}.

Finally we return to the question of SLRs, which provide the
strongest single line of evidence linking formation of the Sun and
Solar System to the environment around a massive star.  While most
Calcium-Aluminum-rich Inclusions (CAIs) require a fairly uniform
abundance of $^{26}$Al/$^{27}$Al of $\sim 5 \times 10^{-5}$ \citep{mdz95}
at the time they formed, others (the so-called ``FUN inclusions,'' for
``Fractionated Unusual Nuclear effects'') contain no \Al.  FUN inclusions
appear to the among the most primitive of the CAIs.  The data are easiest
to understand if there were a single injection of freshly synthesized
material after the disk had formed, during the few $\times 10^5$ year
period over which CAIs were produced \citep{sg98}.  Such events are a
natural consequence of our scenario.  But SLRs are more than just tracers
of the astrophysical environment in which the Sun formed.  Decay of \Al\
was an important source of energy responsible for differentiation of
planetesimals \citep{gm93}, which in turn are believed to be the source
of much of Earth's water \citep{mor00}.  The habitability of Earth may
be directly tied to the larger interstellar environment in which the Sun
formed, and in particular to its location in proximity to a massive star
at the time it went supernova.

It is clear that the astrophysical, meteoritic, and planetary science
communities have a great deal to learn from each other.  Like the proverb
of the Blind Men and the Elephant, we will not have a complete picture
of the formation and evolution of our Sun and Solar System until we have
more completely integrated the insights from all three fields.  While we
are fortunate to have a region of low-mass star formation like
Taurus-Auriga so close at hand, it is important to keep in mind another
proverb about a man who chose to look for his keys under a street lamp
because the light there was better.  In the introduction to
their paper on modes of star formation, \citet{am01} note that, ``If
most star formation takes place within sufficiently dense environments,
then the current theory of star formation could require substantial
modification, or perhaps even a new paradigm.''  It is time for each of
our communities to face more directly the implications of that 
prescient statement.

\acknowledgments
We would like to acknowledge conversations with many colleagues regarding
this work.  We thank John Bally and Alan Boss for their insights and
encouragement regarding this work.  In particular, we acknowledge the
contributions of Laurie Leshin and Kevin Healy, who have played a key
role in this research.  Finally, we owe a special debt of gratitude to
Bo Reipurth for his patience when personal circumstances seriously
delayed the final submission of this manuscript.  This work has been 
supported in part by grant HST-GO-08222.01-A from the Space Telescope 
Science Institute.


\newpage

\end{document}